\documentstyle[12pt]{article}

\makeatletter
\def\appendix{\par\clearpage
  \setcounter{section}{0}
  \setcounter{subsection}{0}
  \@addtoreset{equation}{section}
  \def\@sectname{Appendix~}
  \def\theequation{\thesect.\arabic{equation}}
  \def\thesect{\Alph{section}}
  \def\thesection{\@sectname\Alph{section}}}
\makeatother
\begin{document}

\input{psfig.sty}

\begin{titlepage}
\hskip 12cm \vbox{\hbox{UNICAL-TH 99/1}\hbox{BITP-99-2E}
\hbox{DFPD 99/TH/6}
\hbox{March 1999}}
\vskip 0.3cm
\centerline{\bf FIXED-ANGLE ELASTIC HADRON SCATTERING$^{~\diamond}$}
\vskip 1.0cm{  R. Fiore$^{1\dagger}$, L. L. Jenkovszky$^{2\ddagger}$,
 V. K. Magas$^{3\circ}$, F. Paccanoni$^{4\ast}$}
\vskip .5cm
\centerline{$^{1}$ \sl  Dipartimento di Fisica, Universit\`a della Calabria,}
\centerline{\sl Istituto Nazionale di Fisica Nucleare, Gruppo collegato di
Cosenza}
\centerline{\sl Arcavacata di Rende, I-87030 Cosenza, Italy}
\vskip .2cm
\centerline{$^{2}$ \sl  Bogoliubov Institute for Theoretical Physics,}
\centerline{\sl Academy of Sciences of the Ukraine}
\centerline{\sl 252143 Kiev, Ukraine}
\vskip .2cm
\centerline{$^{3}$ \sl Section of Theoretical Physics (SENTEF)}
 \centerline{\sl Department of Physics, University of Bergen}
\centerline{\sl Allegaten 55, 5007 Bergen, Norway}
\vskip .2cm
\centerline{$^{4}$ \sl  Dipartimento di Fisica, Universit\`a di Padova,}
\centerline{\sl Istituto Nazionale di Fisica Nucleare, Sezione di Padova}
\centerline{\sl via F. Marzolo 8, I-35131 Padova, Italy}
\vskip 0.5cm
\begin{abstract}
The scattering amplitude in the dual model with Mandelstam analyticity 
and trajectory $\alpha (s)=\alpha _{0} -
\gamma \ln \left[ (1+\beta \sqrt{s_{0} -s})/(1+\beta \sqrt{s_{0} } )\right]$
is studied in the limit $s,|t|\rightarrow \infty ,\ s/t=const.$ By using the 
saddle point method, a series decomposition for the scattering amplitude is 
obtained, with the leading and two sub-leading terms calculated explicitly.
\end{abstract}
\vskip .3cm
\hrule
\noindent

\noindent
$^{\diamond}${\it Work supported in part by the Research Council of Norway
(programs for nuclear and particle physics, supercomputing, and free 
projects), in part by the Ministero italiano dell'Universit\`a e della Ricerca 
Scientifica e Tecnologica and in part by the INTAS grant 93-1867-extension}
\vfill
$\begin{array}{ll}
^{\dagger}\mbox{{\it email address:}} &
   \mbox{FIORE~@CS.INFN.IT}
\end{array}
$

$ \begin{array}{ll}
^{\ddagger}\mbox{{\it email address:}} &
 \mbox{JENK~@GLUK.APC.ORG}
\end{array}
$

$ \begin{array}{ll}
^{\circ}\mbox{{\it email address:}} &
 \mbox{MAGAS~@SENTEF2.FI.UIB.NO}
\end{array}
$

$ \begin{array}{ll}
^{\ast}\mbox{{\it email address:}} &
   \mbox{PACCANONI~@PADOVA.INFN.IT}
\end{array}
$
\vfill
\end{titlepage}
\eject
\textheight 210mm
\topmargin 2mm
\baselineskip=24pt

\section{Introduction:  Wide-angle scattering in QCD}

Although attempts to apply perturbative QCD to wide-angle elastic hadron
scatterings have been undertaken in a number of papers [1-10], explicit
predictions have been available only for elastic processes involving external
photons, such as $\gamma + \gamma\rightarrow$ hadrons, Compton scattering
of hadrons, etc.

Predictions based on perturbative QCD rest on three premises: 1) 
hadronic interactions become weak at small invariant separation
$r\ll \Lambda_{QCD}^{-1}$; 2) the perturbative expansion in
$\alpha_s(Q)$ is well-defined; and 3) factorization, implying that all
effects of collinear singularities, confinement, non-perturbative
interactions and bound state dynamics can be isolated at large momentum
transfer in terms of the (process independent) structure functions
$G_{i/H}(x,Q),$  fragmentation functions $D_{H/i}(z,Q)$ or, in the case of
exclusive processes, distribution amplitudes $\Phi_H(s_i,Q).$
Consequently the hadronic scattering amplitude takes the form
\begin{equation}
A=\int\prod_H\phi_H(x_i,Q) T(x_i,p_H;Q)[dx_i]~,
\label{eq1}
\end{equation}
where $\Phi(x_i,Q)$ is a universal distribution amplitude which gives
the probability amplitude for finding the valence $q\bar q$ or $qqq$
in the hadronic wave function collinear up to the scale
$Q=\sqrt{s\over 2}$, and $T$ is the hard scattering amplitude for valence
quark collisions.

The technical complication which has made particularly difficult to
compute the behavior of hadron-hadron amplitudes is the possibility of
multiple scatterings. The standard factorized form for the elastic
scattering of hadrons $\{i\}$ is
\begin{equation}
A_1(s,t)=\int^1_0\prod_{i=1}^4[dx_i]\phi_i(x_i)T(\{x_i\},s,t)~,
\label{eq2}
\end{equation}
where $x_i$ represents collectively the fractional momenta of hadron $i$
carried by its valence partons.

According to this concept, all of the partons collide in a small region of 
the space-time of typical dimension $Q^{-1}$. The relevant contribution to
the amplitude behaves according to the dimensional counting \cite{BF73,
mtm73}, i.e.
\begin{equation}
A_1(s,t)\sim({\mu^2\over s})^{n/2-2}f_1(s/t)
\label{eq3}
\end{equation}
for $n$ partons participating in the hard scattering, $\mu$ representing
hadronic mass scales, which make the amplitude dimensionless.

An extension of this "single-scattering" scenario is the (double) "independent-
scattering" picture, due to P. Landshoff \cite{Lan74}, in which two
pairs of partons
scatter independently off two scattering centers. According to this
picture, the lowest order diagrams contribute with
\begin{equation}
A_m(s,t)\sim\biggl({\mu^2\over s}\biggr)^{(n-m+1)/2-2}f_m(s/t)~,
\label{eq4}
\end{equation}
where $m$ is the number of independent scatterings.
If so, the multiple scattering should dominate in the case of wide angle 
scattering.

A solution to this problem was pointed out in Refs.~\cite{LB80} and 
\cite{LP81}, where it was shown that the Sudakov logarithms associated with
the rescattering diagrams do not cancel. In the leading logarithmic
approximation they exponentiate to suppress the typical double scattering
contribution by a factor
\begin{displaymath} 
exp(-const\ln Q^2\ln(\ln Q^2))~,
\end{displaymath}
characteristic of the Sudakov suppression in QCD.

More quantitatively \cite{M81},
\begin{equation}
A_2\sim{1\over{Q^4}}\biggl({Q\over \Lambda_{QCD}}\biggr)^{1-2c\ln (1/r)}\,,
\label{eq5}
\end{equation}
where
\begin{displaymath}
r=2c/(1+2c)~,~~~\ \ \ \ c=32/(33-2n_f)~,
\end{displaymath}
and $n_f$ is the number of flavors. Interestingly, for $n_f=3$ the
power turns out to be $Q^{-3.8},$ nearly the same as the dimensional
counting power $Q^{-4}$ in the single-scattering scenario.

Higher order diagrams were calculated e.g. in Ref.~\cite{FSZ89}, however soon
it became evident that even the first order QCD correction involves 
an immense number of Feynman diagrams, so further attempts to go
beyond the simple quark counting rule were abandoned.

It may be that perturbative QCD is not the relevant (or not
the only physically interesting) expansion of the wide-angle
scattering amplitude. Recent developments in M-branes (see e.g. 
Ref.~cite{polchinski}) may open new prospects in the realization of a
hypothetical duality between small and large distances (or,
equivalently, large- and small-angle scattering). The search for a
relevant expansion parameter is of crucial importance on this way.

In this paper we are solving an "inverse problem": we use the known
explicit expression of the dual amplitude with Mandelstam
analyticity (DAMA), that has correct wide angle scaling
behavior.  By identifying it with that resulting from the quark
counting rules, we then calculate two sub-leading terms in the
expansion of the known full dual amplitude and study the behavior
of the resulting series.

\section{ Wide angle behavior of the dual amplitude with
Mandelstam analyticity}
\label{sec1}

Wide-angle scaling behavior within the $S-$matrix approach
 was discussed in Ref.~\cite{coon78}, where
by means of a logarithmic Regge trajectory an interpolation from
the ``soft`` Regge behavior to the ``hard`` scaling regime was
suggested. The motivation of the logarithmic trajectory came
from earlier papers \cite{alik75}, where
a class of dual models requiring a logarithmic trajectory was suggested.

The logarithmic asymptotic behavior of the trajectory and the large angle
scaling behavior are uniquely related also in a different class
of dual models, called dual amplitudes with Mandelstam analytics
(DAMA) \cite{jenk79,bcj80}. The link between this class of models in the
scaling
limit and the parton model in the infinite momentum frame was
studied in Ref.~\cite{sch73}. In all those papers only the leading
asymptotic
($s,|t|\rightarrow \infty,\ s/t = const$) term was treated. The results of
different approaches vary in such details as the form of the
scaling violation (normally, logarithmic), the form of the angular
dependence  $f(\theta )$ and the way active quarks are counted.

In this paper we calculate the sub-leading terms in the pre-asymptotic
(larger $s$ and $|t|$) behavior of DAMA. Since the model is realistic
enough in the sense that it satisfies the general requirements
of the theory (see Refs.~\cite{jenk79,bcj80}), we believe that our result is
universal and thus it may be used as a guide e. g. in QCD calculations.

Apart from the leading term, we have explicitly calculated two
more sub-leading terms. Our technique allows further calculations
of still higher orders, but the obtained first three terms of
the series already show a regular trend that may be interpreted
as the expansion in the running coupling constant
 $g(s) \sim 1/\ln s,$ valid at large
$s$ and $|t|$. This situation takes place for anyone trajectory
with the logarithmic asymptotic.

The aim of the present paper is two-fold. First, by identifying the
leading term of the asymptotic (wide-angle) expansion of DAMA with that
derived from perturbative QCD \cite{LB80} we tentatively assume that the
DAMA in the wide angle asymptotic region is equivalent to the
asymptotically free regime in QCD. With this identification in mind, we
calculate within DAMA corrections to the leading term in the hope that their
form may give some insight into the relevant corrections in perturbative
QCD that are known to be very complicated.

Clearly, the above identity has the chance to be true only in the vicinity
of the wide angle region (small distances), where perturbative calculations
are assumed to be still valid.

The second aspect is purely phenomenological. Since, however, the experimental
situation in the wide-angle region did not change for almost two decades,
we are left with the earlier fits to the data.

Let us now calculate the ``perturbative`` expansion of DAMA.
We write the elastic scattering amplitude for spinless particles
in the following symmetric form \cite{bcj80}:
\begin{equation}
 A(s,t,u)=C(s-u)\lbrack D(s,t)-D(u,t)\rbrack~,
\label{eq6}
\end{equation}
where $C$ is a constant and
\begin{equation}
D(s,t)=\int\limits_{0}^{1}dx\left( \frac{x}{g} \right)  ^{-\alpha
(s' )} \left( \frac{1-x}{g} \right) ^{-\alpha (t' )}~.
\label{eq7}
\end{equation}

Here
 $s' =s(1-x),\ \ \ t' =tx$
and $g$ is a dimensionless parameter, $g>1$. Only one, leading trajectory
was included and it was chosen in a simple, but
 representative form:
\begin{equation}
 \alpha (s)=\alpha _{0} -\gamma \ln \left( \frac{1+\beta \sqrt{s_{0} -s}}
{1+\beta \sqrt{s_{0} } } \right)~,
\label{eq8} 
\end{equation}
that account both for the threshold and the asymptotic behavior and
is nearly linear for very small  $|s|,|s|<<s_{0}\,.$
For simplicity we have included only the leading trajectories in both 
channels: the Pomeron trajectory in the $t$-channel and the exotic
trajectory in the $s$-channel. While the parameters of the Pomeron
trajectory are well known, only a little is known about the exotic
trajectory. Fortunately, this has no substantial effect on our results, since
our goal is the functional form of the series and its individual
terms rather than fits to the data. Given the scarcity of the data
and the freedom available in the model, the wide-angle behavior of
DAMA cannot be determined completely.

Let us consider the asymptotic behavior of Eq.~(\ref{eq7}) in the limit
 $s,\left| t\right| \rightarrow \infty ,\ s/t=const.$
For the Regge trajectories we have
\begin{equation}
\alpha(s)=\alpha (0)-\frac{\gamma }{2} \ln \delta ^{2}+ i\pi \frac{\gamma}
{2}- \frac{\gamma }{2}\ln s=-a-\lambda~,
\label{eq9}
\end{equation}
\begin{equation}
\alpha (t)=\alpha (0)-\frac{\gamma }{2} \ln \delta ^{2}-
 \frac{\gamma } {2}\ln\left(-\frac{t}{s}\right)-\frac{\gamma }{2}\ln s=
 -b-\lambda~,
\label{eq10}
\end{equation}
with
\begin{displaymath}
a=-\alpha (0)+\frac{\gamma }{2} \ln \delta ^{2} -i\pi 
\frac{\gamma }{2}~,~~~~
b=-\alpha (0)+\frac{\gamma }{2} \ln \delta ^{2} +\frac{\gamma }{2} 
\ln(-\frac{t}{s})~,
\end{displaymath}
\begin{equation}
\delta=\frac{\beta \sqrt{s_{0}}}{1+\beta \sqrt{s_{0} } }~,~~~~~~
\lambda = \frac{\gamma }{2} \ln s~.
\label{eq11}
\end{equation}
From here on, $s, t, u$ will be dimensionless variables, measured
in units of $s_0$.

In this domain the saddle point method can be used to calculate
the integral in Eq.~(\ref{eq7}) \cite{mag97}. To do this we can rewrite
Eq.~(\ref{eq7}) in the following form 
\begin{equation}
D(s,t)=(2g)^{-a-b-2\lambda} g^{\gamma \ln 2} {1\over 2}
\int\limits_{-1}^{1}g(u)~e^{\lambda f(u)} du~,
\label{eq12}
\end{equation}
where we have changed the variable $x$ to $u$, $x=(1-u)/2$, and introduced 
new functions:
\begin{equation}
g(u)=(1-u)^{\tilde{a}}(1+u)^{\tilde{b}}e^{\gamma \ln{\frac{1-u}{2}}
\ln{\frac{1+u}{2}}}~,
\label{eq13}
\end{equation}
\begin{equation}
f(u)=\ln (1-u^2) ~,
\label{eq14}
\end{equation}
\begin{equation}
\tilde{a}=a-\frac{\gamma}{2}\ln g ~,\ \ \ \ \ \tilde{b}=b-\frac{\gamma}{2}\ln g~.
\label{eq15}
\end{equation}
We see now that $f(u)$ has a sharp maximum at the saddle point $u_0=0$.

We quote the explicit expression for the saddle point expansion in the 
\ref{ap_coef}.
Using formulas from this Appendix 
we obtain the power series for $D(s,t)$ in \ref{powerD}. It reads
\begin{equation}
D(s,t)\approx \frac{A_1 s^{-\gamma \ln 2g}}{\sqrt{\gamma \ln s}} 
\left(-\frac{t}{s}\right)^{-\frac{\gamma }{2} \ln 2g}\left\{1+
\frac{h_1(\tilde{a},\tilde{b})}{\gamma \ln s}+\frac{h_2(\tilde{a},\tilde{b})}{(\gamma \ln s)^2}\right\}~,
\label{eq16}
\end{equation}
where $A_1,\ h_1, \ h_2$ are given by the expressions
(\ref{b5}, \ref{b8}, \ref{b9}).
The expression for $D(u,t)$ can be calculated in a similar way (see 
Eq.~(\ref{b10}) in \ref{powerD}).

In the kinematical region
 $s, \left| t\right| \rightarrow \infty,\ t/s=const$
we can use the  substitutions
\begin{equation}
t\approx -s\cdot \sin ^{2} \left( \theta/2 \right)~,~~~~~~
u\approx -s\cdot \cos ^{2} \left( \theta/2 \right)~.
\label{eq17}
\end{equation}

Substituting the results for $D(s,t)$ and $D(u,t)$ into Eq.~(\ref{eq6}) and 
changing the variables we get the expression
for the full amplitude as a function of the $s$ and $\theta$ variables 
(see Eq.~(\ref{b14}) in \ref{powerD}):
\begin{equation}
A(s,\theta)\approx \frac{C A s^{-N}}{\sqrt{\gamma \ln s}} f(\theta)
I(s,\theta )~,
\label{eq18}
\end{equation}
where $A,\ N,\ f(\theta),\ I(s,\theta )$ are given by the expressions
(\ref{b13}, \ref{b15} - \ref{b17}).

To summarize, we have expanded the wide-angle scattering amplitude in a 
power series of $1/\ln{s}$ and have evaluated explicitly the
coefficients of the first two terms (beyond the leading one).

\section{Comparison with the data and discussion of the results}
\label{sec2}

\begin{figure}[htb]
\centerline{\psfig{figure={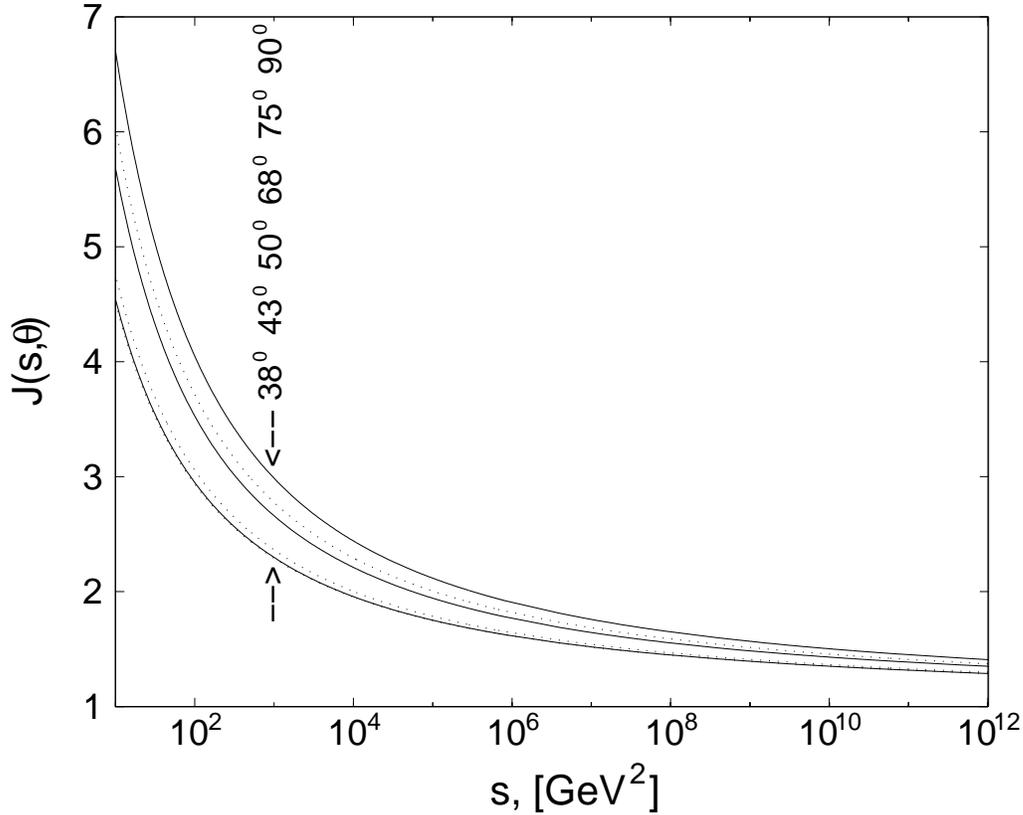},height=11.0cm}}
\caption{\footnotesize Cross section $\frac{d\sigma }{dt}$ for {\bf pp}
$\rightarrow$ {\bf pp} scattering
at various center of mass scattering angles. Both axes are in logarithmic 
scale. Stars denote the experimental points from Ref.~\cite{LP73}. 
The straight lines correspond to a falloff of 
$\sim 1/s^{10}$. They are calculated according to the power series for the 
scattering amplitude, discussed above ($\frac {d\sigma}{dt}=
 \frac{4\pi}{(s s_0)^2}|A(s,\theta)|^2$), with the following set of 
parameters: 
$\alpha_0 = 1$,
$N=4,\ \gamma = 2.84 \ (g=2.9),$  $\beta=0.05 \ GeV^{-1}$, $C=2.7\cdot10^{-14}
 GeV^{-2}$ and $s_0 = 4m_\pi^2$.
}
\label{fig1}
\end{figure}

New experimental data on wide-angle scatterings are not likely to
appear any more because of the simple reason that as energy increases more
particles tend to fly in the forward direction and there is no chance
to detect e.g. the proton-proton differential cross section at
$90^{\circ}$ for, say, $\sqrt s> 10\ GeV$. ``Wide angles'', of course,
extend beyond $90^{\circ}$.  Still the complication due to the huge number
of Born diagrams contributing to large angle exclusive reactions \cite{LB80}, 
overwhelming the contribution due to the Landshoff pinch singularity \cite{Lan74},
will remain for long topical in this field.  We use the data given in the
compilation of \cite{LP73} to fix the scale. The errors, quoted in the
original papers (see Ref.~\cite{LP73} and references therein), are
typically about 10 
overall normalization factor, the "quark counting power" in the cross
section being set equal to $N=4$ in the case of proton-proton
cross section, in agreement with the data \cite{LP73, LB80} 
(see Fig.~\ref{fig1}).


\begin{figure}[htb]
\centerline{\psfig{figure={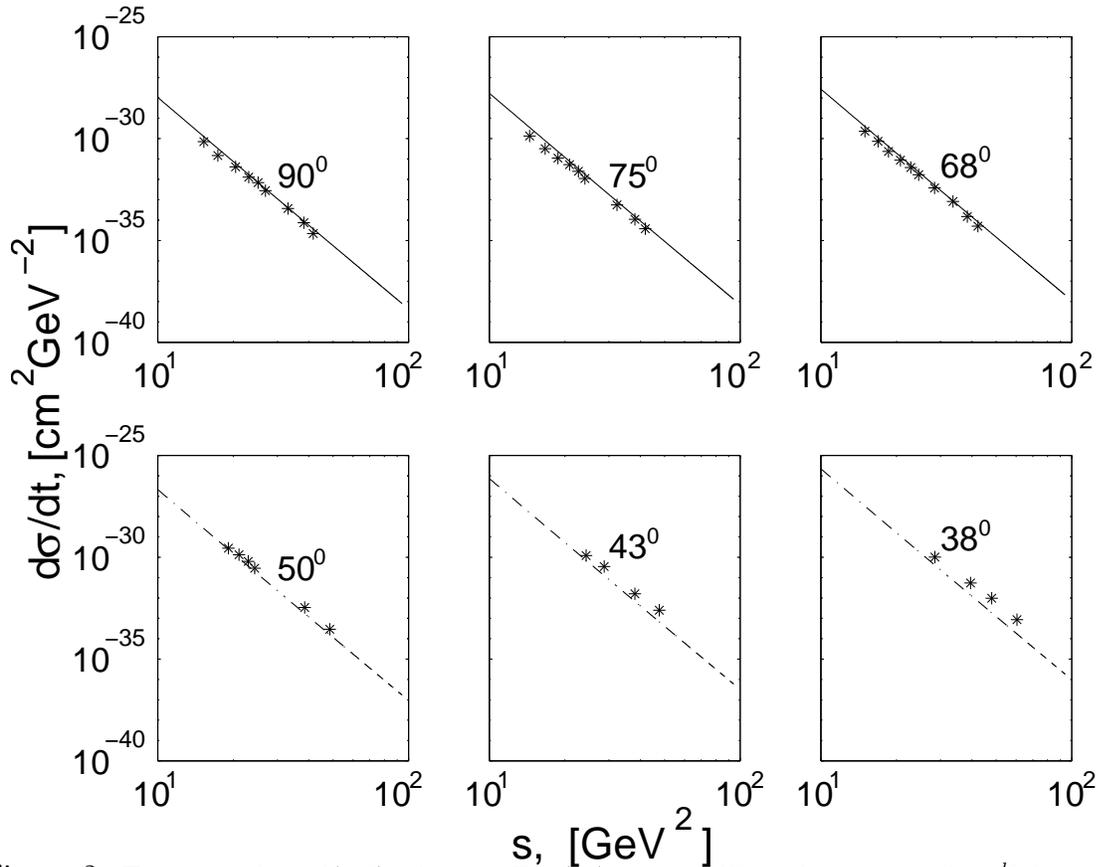},height=11.0cm}}
\caption{\footnotesize
 The corrections  $J(s,\theta)$, given by
Eq.~(\ref{eq19}), to the differential cross section  $\frac{d\sigma }{dt}$
for {\bf pp} $\rightarrow$ {\bf pp} scattering .
We have used the same values of parameters as in Fig.~\ref{fig1}: 
$\alpha_0 = 1$, $N=4,\ \gamma = 2.84\ (g=2.9),$ $\beta=0.05 \ GeV^{-1}$
 and $s_0 = 4m_\pi^2$, coming from the comparison with  the data.}

\label{fig2}
\end{figure}

Our main goal is the behavior of the scaling-violating corrections
to the leading term obeying quark counting rules. Fig. \ref{fig2} shows the
relative contribution of these terms. 
We draw the correction power series:
\begin{displaymath}
J(s,\theta)=|I(s,\theta)|^2 \approx 1+ 2\frac{Re(f_1(\theta)/Z(\theta))}
{\gamma \ln s}+
2\frac{Re(f_2(\theta)/Z(\theta))}{(\gamma \ln s)^2}+\frac{|f_1(\theta)
/Z(\theta)|^2}{(\gamma \ln s)^2}
\end{displaymath}
\begin{equation}
+O({1\over \lambda^{3}})~,
\label{eq19}
\end{equation}
where $f_1(\theta),~f_2(\theta),~Z(\theta)$ are given by expressions
 (\ref{b18} - \ref{b20}).
We can see that the corrections
are quite large for small $s$, especially for angles close to $90^0$. That is 
not a surprise, since the lowest order of our
expansion is valid for large $s$ ($\gamma \ln s /2 >> 1$). In the experimental 
energy interval the corrections give factor $4-6$ to the cross sections and 
should not be neglected. This was missed in the references \cite{jenk79, bcj80}. 
 Moreover we find that the 
corrections are very sensitive to variations of $\beta$ and $\gamma$.


\vskip 1cm
{\bf Acknowledgment}
V.K. Magas is thankful for the hospitality extended to him by the Bogolyubov 
Institute for Theoretical Physics in Kiev where part of this work was done.

\appendix

\section{$\!\!\!\!\!\!$: Coefficients in the saddle point method}
\label{ap_coef}

In this Appendix we present the explicit expression for the saddle point 
expansion from Ref.~\cite{fed87}.
Here  $f^{(k)}(u_{0})\equiv f _{k}~,~~~g^{(k)}(u_{0})\equiv g _{k}~.$
\begin{equation}
\int\limits_{-1}^{1}g(u)~e^{\lambda f(u)} du = e^{\lambda f(u_0)}
\sqrt{\frac{\pi}{\lambda}}\left[a_0+{a_1\over \lambda}+{a_2\over \lambda^2}+
O(\frac{1}{\lambda^3})\right]~.
\label{a1}
\end{equation}
where
\begin{displaymath}
a_0=\psi_1g_0~,~~~a_1={1\over 4}\left[g_2\psi_1^3+3g_1\psi_1\psi_2+
g_0\psi_3\right]~,
\end{displaymath}
\begin{equation}
a_2={1\over32}\left[g_4\psi_1^5+10g_3\psi_1^3\psi_2+10g_2\psi_1^2\psi_3+
15g_2\psi_2^2\psi_1+5g_1\psi_4\psi_1+10g_1\psi_3\psi_2+g_0\psi_5\right]~.
\label{a2}
\end{equation}
\begin{displaymath}
\psi_1=\sqrt{\frac{2}{-f_2}}~,~~~\psi_2=-{1\over 3}f_3f_2^{-1}\psi_1^2~,~~~
\psi_3=\left[-{1\over 4}f_4f_2^{-1}+{5\over 12}f_3^2f_2^{-2}\right]\psi_1^3~,
\end{displaymath}
\begin{displaymath}
\psi_4=\left[-{1\over 5}f_5f_2^{-1}+f_4f_3f_2^{-2}-{8\over 9}f_3^3f_2^{-3}\right]
\psi_1^4~,
\end{displaymath}
\begin{equation}
\psi_5=\left[-{1\over 6}f_6f_2^{-1}+{7\over 6}f_5f_3f_2^{-2}+{35\over 48}f_4^2
f_2^{-2}+{385\over 144}f_3^4f_2^{-4}-{35\over 8}f_4f_3^2f_2^{-3}\right]\psi_1^5
~.
\label{a3}
\end{equation}

\section{$\!\!\!\!\!\!$: Calculations of the scattering amplitude}
\label{powerD}
Using the definitions of functions $g(u)$, $f(u)$ (\ref{eq13}, \ref{eq14}) 
we obtain
\begin{equation}
f_2=-2~,~~~f_3=0~,~~~f_4=-12~,~~~f_5=0~,~~~f_6=-240~,
\label{b1}
\end{equation}
\begin{displaymath}
g_0=e^{\gamma \ln^2 2}~,~~~g_2=g_0\left[
(\tilde{a}-\tilde{b})^2-(\tilde{a}+\tilde{b})+2\gamma(\ln 2-1)\right]~,
\end{displaymath}
\begin{displaymath}
g_4=g_0\left[\tilde{a}(\tilde{a}-1)(\tilde{a}-2)(\tilde{a}-3)-4\tilde{a}(\tilde{a}-1)(\tilde{a}-2)\tilde{b}+
6\tilde{a}(\tilde{a}-1)\tilde{b}(\tilde{b}-1)-4\tilde{a}\tilde{b}(\tilde{b}-1)
(\tilde{b}-2)+\right.
\end{displaymath}
\begin{displaymath}
\ \left. \tilde{b}(\tilde{b}-1)(\tilde{b}-2)(\tilde{b}-3)+
12\gamma\left((\tilde{a}-\tilde{b})^2-(\tilde{a}+\tilde{b})
\right)(\ln 2-1)+12\gamma^2(\ln 2-1)^2\right.
\end{displaymath}
\begin{equation}
\left.+ 2\gamma(6\ln 2-5)\right]~.
\label{b2}
\end{equation}

From Eq. (\ref{a3}, \ref{b1}) we get
\begin{equation}
\psi_1=1~,~~~\psi_2=0~,~~~\psi_3=-{3\over 2}~,~~~\psi_4=0~,~~~
\psi_5={25\over 4}~.
\label{b3}
\end{equation}

Finally we get
\begin{equation}
D(s,t)\approx \frac{A_1 s^{-N_1}}{\sqrt{\gamma \ln s}} 
\left(-\frac{t}{s}\right)^{-\frac{\gamma }{2} \ln 2g}I(\tilde{a},\tilde{b},s)~,
\label{b4}
\end{equation}
where
\begin{equation}
A_1=(2g)^{2\alpha_0-\gamma \ln \delta ^{2} +\gamma \ln 2+i\pi
\gamma /2}\sqrt{\frac{\pi }{2} }~,
\label{b5}
\end{equation}
\begin{equation}
N_1=\gamma \ln 2g~,
\label{b6}
\end{equation}
\begin{equation}
I(\tilde{a},\tilde{b},s)=\left\{1+\frac{h_1(\tilde{a},\tilde{b})}{\gamma \ln s}+
\frac{h_2(\tilde{a},\tilde{b})}{(\gamma \ln s)^2}\right\}~,
\label{b7}
\end{equation}
Coefficients $h_1(\tilde{a},\tilde{b}),~h_2(\tilde{a},\tilde{b})$ 
are calculated from 
(\ref{a2}, \ref{b2}, \ref{b3})
\begin{equation}
h_1(\tilde{a},\tilde{b})= -\left(\frac{3}{4}-\frac{g_2}{2g_0}\right)~,
\label{b8}
\end{equation}
\begin{equation}
h_2(\tilde{a},\tilde{b})= \left({25\over 32} + \frac{g_4}{8g_0}-\frac{15g_2}{8g_0}\right)~.
\label{b9}
\end{equation}

The expression for $D(u,t)$ can be calculated in a similar way. It turns
out to be
\begin{equation}
D(u,t)\approx \frac{A_2 s^{-N_1}}{\sqrt{\gamma \ln s}}
\left(\frac{ut}{s^{2}}\right)^{-\frac{\gamma }{2} \ln 2g}I(\tilde{c},\tilde{b},s)~,
\label{b10}
\end{equation}
where 
\begin{displaymath}
A_2=(2g)^{2\alpha_0-\gamma \ln \delta ^{2} +\gamma \ln 2}
\sqrt{\frac{\pi }{2} }~,
\end{displaymath}
\begin{equation}
\tilde{c}=c-{\gamma\over2}\ln g=-\alpha(0)+\frac{\gamma }{2} \ln \delta ^{2} +\frac{\gamma }{2}
\ln \left(-\frac{u}{s}\right)-{\gamma\over2}\ln g~.
\label{b11}
\end{equation}

Substituting Eqs.~(\ref{b4}) and (\ref{b10}) into Eq.~(\ref{eq6}) we get 
the expression for the full amplitude:
\pagebreak
\begin{displaymath}
A(s,t,u)\approx C \frac{A}{s_0} \frac{s^{-N_1}} {\sqrt{\gamma \ln s}}\
(s-u) s_0
\left\{ (2g)^{\frac{i\pi \gamma }{2} } \left(-\frac{t}{s}\right)^
{-\frac{\gamma }{2} \ln 2g} I(\tilde{a},\tilde{b},s) \right. 
\end{displaymath}
\begin{equation}
\left. \hfill -
 \left(\frac{tu}{s^{2} }\right)^{-\frac{\gamma }{2} \ln 2g}I(\tilde{c},\tilde{b},s)\right\}
 \hfill ~,
\label{b12}
\end{equation}
where
\begin{equation}
A=(2g)^{2\alpha (0)-\gamma
\ln \delta ^{2} +\gamma \ln 2}\sqrt{\frac{\pi }{2} } s_0~,
\label{b13}
\end{equation}

In the kinematical region
 $s, \left| t\right| \rightarrow \infty, \ t/s=const$
we can use the  substitutions (\ref{eq17}).
So, the expression for the scattering amplitude as a function of $s$ and 
$\theta$ appears to be
\begin{equation}
A(s,\theta )\approx C \frac{ A s ^{-N}} {\sqrt{\gamma \ln s}}
f(\theta) I(s,\theta )~,
\label{b14}
\end{equation}
where
\begin{equation}
N=N_1-1=\gamma \ln 2g -1~,
\label{b15}
\end{equation}
\begin{equation}
f(\theta)=\left(1+\cos ^{2} \frac{\theta }{2} \right)\left(\sin
\frac{\theta }{2} \right) ^{-\gamma \ln 2g} Z(\theta)~,
\label{b16}
\end{equation}
\begin{equation}
I(s,\theta )=1+\frac{f_1(\theta)}{Z(\theta) \gamma \ln s}
+\frac{f_2(\theta)}{Z(\theta) (\gamma \ln s)^2}~,
\label{b17}
\end{equation}
\begin{equation}
f_1(\theta)= h_1(\tilde{a},\tilde{b})
(2g)^{\frac{i\pi \gamma }{2} } - 
h_1(\tilde{c},\tilde{b})(2g)^{-\gamma \ln\cos\frac{\theta}{2}}~,
\label{b18}
\end{equation}
\begin{equation}
f_2(\theta)= h_2(\tilde{a},\tilde{b})
(2g)^{\frac{i\pi \gamma }{2} } - 
h_2(\tilde{c},\tilde{b})(2g)^{-\gamma \ln\cos\frac{\theta}{2}}~,
\label{b19}
\end{equation}
\begin{equation}
Z\left(\theta\right)=(2g)^{\frac{i\pi \gamma }{2} }
-(2g)^{-\gamma \ln\cos\frac{\theta}{2}}~,
\label{b20}
\end{equation}
\begin{equation}
\tilde{b}=-\alpha (0)+\frac{\gamma }{2} \ln \delta ^{2} +\frac{\gamma }{2} \ln
\left( \sin ^{2} \left( \frac{\theta }{2} \right)\right)-{\gamma\over 2}\ln g~,
\label{b21}
\end{equation}
\begin{equation}
\tilde{c}=-\alpha(0)+\frac{\gamma }{2} \ln \delta ^{2} +\frac{\gamma }{2}
\ln \left(\cos ^{2} \left( \frac{\theta }{2} \right)\right)
-{\gamma\over 2}\ln g~.
\label{b22}
\end{equation}

\vfill \eject

\end{document}